# Voltammetric Determination of Paraquat Using Graphite Pencil Electrode Modified with Doped Polypyrrole

Maryam Sayyahmanesh[1,*], Sara Asgari[2], Azam S. Emami Meibodi[2], Taha Mohseni Ahooyi[3]

**Abstract**
Recognition and determination of paraquat (PQ) using graphite pencil electrode (GPE) modified with polypyrrole (Ppy) doped with Eriochrome blue-black B (EBB) is reported. To that end, a thin film of Ppy was deposited onto the electrode surface by electropolymerization in the presence of a functional doping ion, EBB. The Ppy/EBB-coated electrode was templated by PQ ion and then the performance of the molecularly imprinted EBB/Ppy/GPE was evaluated by voltammetric technique. The prepared electrode exhibited considerable increase in electroactivity of the sensor toward this herbicide compared to the non-imprinted electrode. To enhance the detection capability of the prepared system, the factors controlling its response were investigated and optimized using differential pulse voltammetry. The proposed analytical procedure was proved to be applicable in the concentration range of 5 to 50 µM ($R^2$ = 0.9939) and detection limit of (3σ) 0.22 µM. Ultimately, the proposed analytical methodology was applied to ascertain possible interferences and investigate real samples of dam water. Recovery factors were achieved between the range of 93-104%.

## 1. Introduction

Paraquat (1,1´-dimethyl-4,4´-bipyridinium dichloride) is a toxic chemical that is extensively used as an herbicide, primarily for weed and grass control [1]. Hence, many farmers rely on this herbicide to catch up the growing demand of crops. However, accidental or occupational ingestion or inhalation of this pesticide could be seriously poisoning. US Environmental Protection Agency (EPA) classifies PQ as "restricted use" and limits its accessibility to only licensed consumers [2]. As PQ residues are naturally non-biodegradable and can be easily mixed with food, water, or other beverages, several methods have been proposed to measure the PQ contamination levels [3]. Over the past years, the electro-analytical measurement technique has received an increasing attention from the research community since it is relatively inexpensive and quickly implementable. In the late 1960 Brand and Fleet [4], for the first time, reported the use of mercury coated platinum wires as voltammetric sensors in order to detect dithiocarbomate pesticides. Solid-phase extraction [5-8], chromatographic determination [9-11] and various catalytic biosensors [12, 13] are some of the methods described erstwhile to detect and drive out various components. Molecularly imprinted polymer-based electrode is a modified electrode for environmental analysis that has become a growing field over the last two decades. Generally, the procedure of synthesizing this advanced sensor comprises two steps; i) preparing a compound solution of a template (or target) and functional electroactive monomers by forming non-covalent bonds, and ii) polymerization of the ready monomers around the template molecules with the aid of an appropriate dopant, if needed, and an initiator agent. After removing the template molecules, the available cavities are created that lead to conformational recognition of the polymer matrix, acting as intelligent sites complementary to the target analyte in complex specimens [14]. Organic synthetic conductors such as poly aniline, thiophene, and pyrrole, known as synthetic metals, have been of particular research interest in the fabrication of molecularly imprinted electrodes [15-21]. Among different conductors, pyrrole has attracted more attention because of its fibrous structure, biocompatibility, the ease of preparation, high stability in air and aqueous media [22]. Because of these favorable features, it is broadly used in electronic capacitors [23], energy storage cells [24], electrochromic devices [25] and electronic chemical sensors. In fabricating electroanalytical sensors, it was applied to modify electrode surfaces in order to determine anionic, cationic and neutral analytes (targets) [26-28]. To prepare ion selective electrode, pyrrole is polymerized in the presence of a counter ion/dopant anion such as chlorate, perchlorate, nitrate, etc., which causes a balanced oxidized state of polypyrrole (Ppy) consisting of poly radical cations. By doing so, these dopant agents are accommodated electrostatically into the polymer chain and make the polymer backbone electroneutral [29, 30]. Film morphology, electrical conductivity, thermal and mechanical stability of the prepared polymer is highly dependent on the structure of the dopant [31].

---
[1],* Corresponding Author, Department of Chemical and Petroleum Engineering, Sharif University of Technology, Tehran, Iran, sayyahmanesh@alum.sharif.ir
[2] Department of Chemistry, Karaj branch, Isamic Azad University, Karaj, Iran.
[3] Chemical and Biological Engineering Department, Drexel University, Philadelphia, PA 19104, USA, tm495@drexel.edu

Overoxidation process is induced on the Ppy film at positive potential, as the overoxidized state of Ppy applied primarily to anionic species. Although this process resulting in loss of polymer electroactivity, electron-rich functional groups of hydroxyl and carboxyl introduced to the pyrrole monomers can attract the cationic species of the electrolyte. Cationic perm-selectivity of the overoxidized polymer film of the Ppy-based electrodes corresponds to the loss of cationic charge of the polymer backbone under this process [32, 33]. Multiple studies have been devoted to the application of the molecularly imprinted overoxidised Ppy to assay various types of pollutant [34-36]. Additionally, adopting more massive dopants with low mobility instead of aforementioned small ones leads to create the configuration of overoxidesied Ppy in order to electroanalytically determine cations. Indeed, small anion counter ions are expelled from the polymer matrix during the reduction process, since the polymer backbone turning in its neutral posture. However, the polymer film neutrality can be preserved during the electrochemical process if the bulky anodic dopants permeate the polymer chain. As such, this conformation can be maintained even if it is subjected to a reduction process [37]. Several research studies have taken advantage of this method to develop Ppy-based electrodes to be sensitive toward cathodic species [27, 30].

In present work, the benefit of using Ppy/EBB/GPE is introduced to voltammetric recognition of herbicide paraquat. After characterization of the modified electrode by scanning electron microscopy (SEM), the parameters that can directly overshadow the system's response were adjusted according to the best conditions under which the procedure could work. Subsequently, this analytical device was examined for the realistic dam water samples.

## 2. Experimental
### 2.1. Apparatus and chemicals

Electrochemical analysis was performed using Autolab potentiostat/galvanostat PGSTAT-30 model (Metrohm) equipped with GPES software interface. A triple electrode system was applied which consisted of a Ppy/EBB/GPE, Ag/AgCl (3M, KCl), and platinum wire as the working electrode, reference electrode and counter electrode, respectively.

To set out measurements pH-7110 model Inolab pH-Ion meter and to characterize electrode surface, scanning electron micrograph AIS2300 Seron model were utilized.

The following chemicals were used during the course of experiments: methyl viologen dichloride hydrate (98%), potassium chloride (99%), sodium sulphate (99%), sodium nitrate (99%), Eriochrome Blue-black B and pyrrole (98%) which was distilled, purified and was kept at 4°C. All chemicals were purchased from Sigma-Aldrich (Schnelldorf, Germany).

### 2.2. Fabrication of imprinted and non-imprinted Ppy/EBB/GPE

A graphite pencil-electrode (PFG) (4mm diameter) whose lateral surface entirely insulated with epoxy resin, attached to an electric wire from one end, was used. At first, its base was polished to get mirror-like surface with 0.05 μm alumina slurry and an emery paper. Then, it was ultrasonicated for 3 min and finally, rinsed with distilled water thoroughly and dried at room temperature. Electropolymerization process was performed by the PFG immersed into the electrolyte solution with 0.05 M pyrrole monomer and 0.1 M EBB at a constant voltage of 0.75 V(vs. Ag/AgCl) for 10 min with the aid of the potentiometric technique and was then purged with $N_2$ [27, 30]. When electropolymerization was done, the PFG was rinsed and then immerserd into the solution containing 0.009 M PQ. After elution of the electrode, consecutive reduction (-0.4 V) /oxidation (-0.5 V) steps were induced in the electrode. Hence, the positively-charged polymer backbone which had become neutral during the reduction process, is negatively-charged by incorporating an anionic dopant agent. As such, appropriate conditions under which PQ could permeate into the polymer matrix were prepared. As a result, a thin film of Ppy/EBB was deposited onto the electrode surface with embedded PQ. The polymer film templated by PQ was driven out to leave a complementary mold on the electrode surface that possesses an appropriate functionality and shape compatible to the target analyte. To that end, the imprinted polymer was subjected to an oxidation step for 5 min to remove PQ template from the polymer chain. Non-imprinted Ppy/EBB/GPE was fabricated as described above just as far as a film of EBB/Ppy formed by electropolymerization onto the GPE surface. Consequently, the electrode was prepared to be applied for further analysis by voltammetric method. The entire process is shown in Fig. 1.

## 2.3. Analytical procedure

Differential pulse voltammetry was performed in a conventional three electrode system in the potential range between -0.8 to 0.8 V. To select the preferable supporting electrolyte with the highest analytical signal, some of which were tested ($Na_2SO_4$, $NaNO_3$ and KCl, all 0.2 M). The best electrochemical response was that of the KCl solution because of its better current responses (results not shown). Hence, it was adopted as a supporting solution accompanying with the $Na_2HPO_4$ buffer (pH 7.0) at a fixed concentration of 0.2 M in all experimental steps. The electrolytic solution was purged with $N_2$ throughout the experimental tests in order to somewhat prevent the unfavorable reaction of the reduced PQ and oxygen [38]. The considered instrumental parameters are as follows: Modulation amplitude of 25 mV, modulation time of 30 ms and step potential of 9 mV. All measurements were conducted at room temperature.

## 3. Result and discussion
### 3.1. Structure characterization

The structural morphology of the modified electrode was analyzed by scanning electron microscopy. Fig 2.a, b and c display surface images of the GPE, non-imprinted and imprinted by Ppy/EBB/GPE, respectively which were taken at the same magnitude. Fig 1.a shows a polished graphite electrode that originally has a rough and dark surface. In the next step, Ppy was deposited singly onto the electrode surface. Fig 1.b illustrates non-imprinted Ppy/EBB/GPE whose surface is more monotonic compared to that of GPE. Fig. 1c is associated with the film of Ppy templated by PQ and layer-embedded onto the GPE after the elution step. As can be seen from this figure, lots of pores have been formed almost uniformly, indicating PQ molecular footprints after extracting the template molecules by washing out from the polymer film onto electrode surface. Fig 1.d shows the image of imprinted Ppy/EBB/GPE with 3μm dimensions and 30k magnitude before washing out the template molecules. According to this image, diffusion and distribution of the ordered template molecules imprinted through the polymer film is evident.

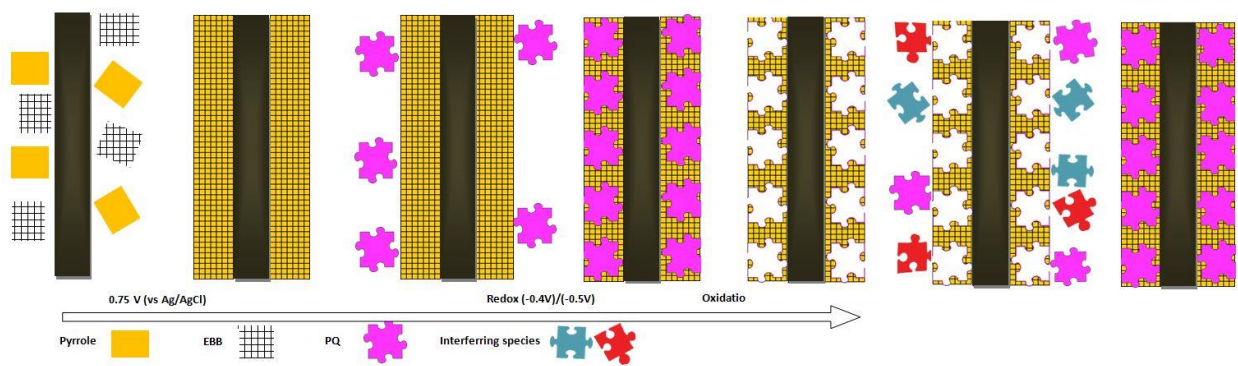

**Fig. 1** Schematic illustration of the fabrication of imprinted Ppy/EBB/GPE

### 3.2. Electrochemical behavior assay of the modified electrode

Prior to developing the protocol for determination of the pesticide, experiments were evaluated using cyclic voltammetry to compare the electrochemical behavior of PQ at the imprinted and non-imprinted Ppy/EBB/GPE (Fig 3). As expected, no voltammetric peak was observed at the non-imprinted Ppy/EBB/GPE for PQ over the potential range while in the case of the imprinted Ppy/EBB/GPE exhibited two cathodic peaks: In the direction of the negative potential values, paraquat cation, $PQ^{2+}$, was incurred a fast and reversible reduction to the radical $PQ^{+}\bullet$ at approximately $-0.7$ V. This radical cation adsorbs onto the electrode surface and upon scanning the potential to more negative values, it was reversibly reduced to the neutral species $PQ^0$ at $-1.07$ V. In other word, $PQ^{2+} + e^- \leftrightarrow PQ^+$; $E_1 = -0.7\,V$ and a quasi-reversible process of $(PQ^+ + e^- \leftrightarrow PQ^0$; $E_2 = -1.07)$ to produce the neutral form of PQ versus Ag/AgCl, which were in agreement with the previous reported studies [39-42].

Subsequently, electroanalysis was evaluated based on the first peak because it presented better signal current response compared to the other one.

### 3.3. Optimization of the experimental parameters

In order to elaborate the sensitive, selective and reliable electrochemical procedure for the quantification of PQ, the main parameters that could directly influence the performance of the sensor were optimized. Therefore, optimization of pH, template concentration and the concentration ratio of the monomer to template were conducted using differential pulse voltammetry method. In this way these smart sensors could be safely used in the industrial process control.

### 3.3.1. pH optimization

Varying pH of the electrolyte solution not only has significant influence on the current response of PQ but also has considerable impact on the stability of the polymer film and the adsorption of PQ on substrate [43, 44]. The effect of pH (2~9) on the potential response was studied at the constant PQ concentration of 30 mM (results not shown). Current signal increased in response to the increase of pH up to 7. As soon as pH exceeds a value above 7, the current response degraded dramatically due to the high concentration of $H^+$ in acidic condition competing with the cation state of PQ to being reduced. On the other hand, the great amount of negative charge in high pH condition causes deactivation of PQ electrochemically. Accordingly, pH of the solution in which the best current response received was 7.

### 3.3.2. Template concentration

Template molecules that are supposed to be accommodated into the polymer matrix during its electrodeposition onto the electrode surface, has profound impact on the effectiveness of the sensor. Template concentration can effect on selectivity, sensitivity capability and intrinsic polymer conductivity. Thus, to clarify the optimal amount of the template concentration of PQ giving rise to the best current signal is necessary. The responses of the imprinted Ppy/EBB/GPE were shown at the fix concentration of monomer (50 mM) with variable concentration of PQ from 1 to 13 mM under the assessed optimized circumstances above. Fig. 4 indicates that as the PQ concentration increases up to 9 mM, the current response gets higher. It can be interpreted that at this concentration pores have been well-formed and the permeability of template molecules into the polymer film increased. With exceeding the concentration of 9 mM, the current responses come to decrease so that it would have an adverse effect on the sensor selectivity. As a result, the best template concentration for this advanced electrode is considered 9 mM.

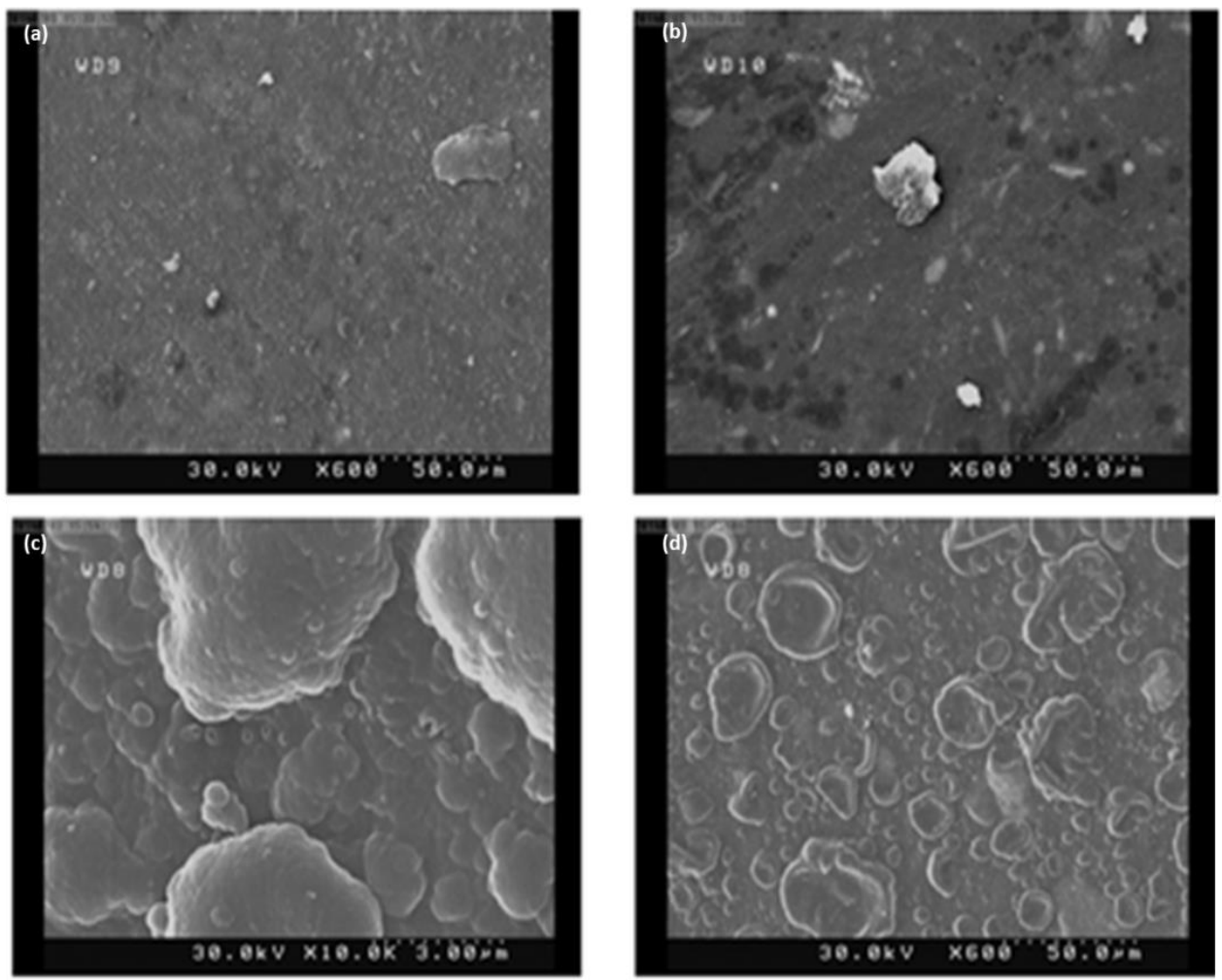

**Fig 2.** SEM micrographs of: (a) GPE, (b) non-templated, (c) templated by Ppy/EBB/GPE, (d) templated by Ppy/EBB/GPE before extracting the template

### 3.3.3. Optimization of the concentration ratio of monomer to template

One of the main variables that have a pronounced effect on the performance of the MICP-modified electrode is the functional monomer to template concentration. To evaluate the effect of monomer concentration, the polymer film was propagated in solutions containing a constant concentration of template, 9 ml of PQ, and different concentrations of pyrrole in the range of 10 to 70 mM Fig 5. The results showed that the current response ascends with increasing pyrrol concentration up to 50mM. It suggests that in this concentration the most interaction between monomer and PQ occurs, additionally, the cavities related to the template molecules were best formed throughout the polymeric film. Thus, in higher or lower concentrations of 50mM these interactions descend due to the alteration of the film thickness, as a result of which the sensitivity toward the analyte (target) diminishes. Therefore, the best monomer to template concentration ratio was estimated to be 6:1.

### 3.4. Efficiency of the imprinted sensor
### 3.4.1. Detection limit and repeatability

The calibration data was carried out under the optimal conditions which specified and discussed above. To that end, certain solutions of PQ were prepared and analyzed at GPE/Ppy MICP using DPV for quantitative evaluation due to its higher sensitivity compared to the cyclic method. The responses were recorded and then the calibration curve was obtained. Based on Fig. 6 the (calibration) peak reduction profile versus the PQ concentration was plotted with dependence on the first peak ($PQ_1$). A highly linear relation between PQ concentration $[PQ]$ and the current signal over the range of 5 to 50 µM was achieved with a fitted equation of

$$\Delta i (\mu A) = 0.0647 [PQ](\mu M) + 0.2167$$
$$(R^2 = 0.9939)$$

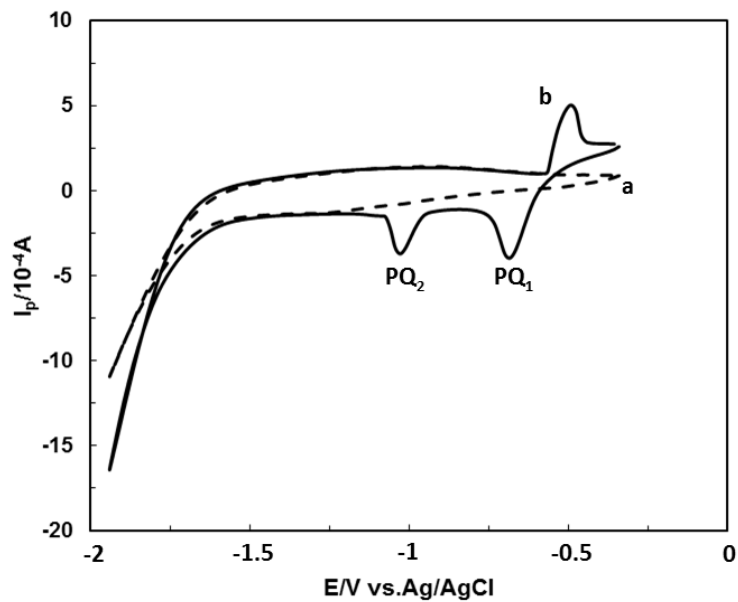

**Fig 3.** Cyclic voltammograms of $9 \times 10^{-3}$ M PQ at (a) non-imprinted Ppy/EBB/GPE and (b) imprinted-Ppy/EBB/GPE in phosphate buffer at pH 7.0 with 0.2 M KCl.

With detection limit (3σ) of $2.2 \times 10^{-7}$ and quantification limit (10σ) of $7.2 \times 10^{-7}$. Also, the proposed electrode was examined the repeatability from 5 replicate measurements with the same PQ concentration of 5 µM; the relative standard deviation was found 3.51% confirming the results are repeatable.

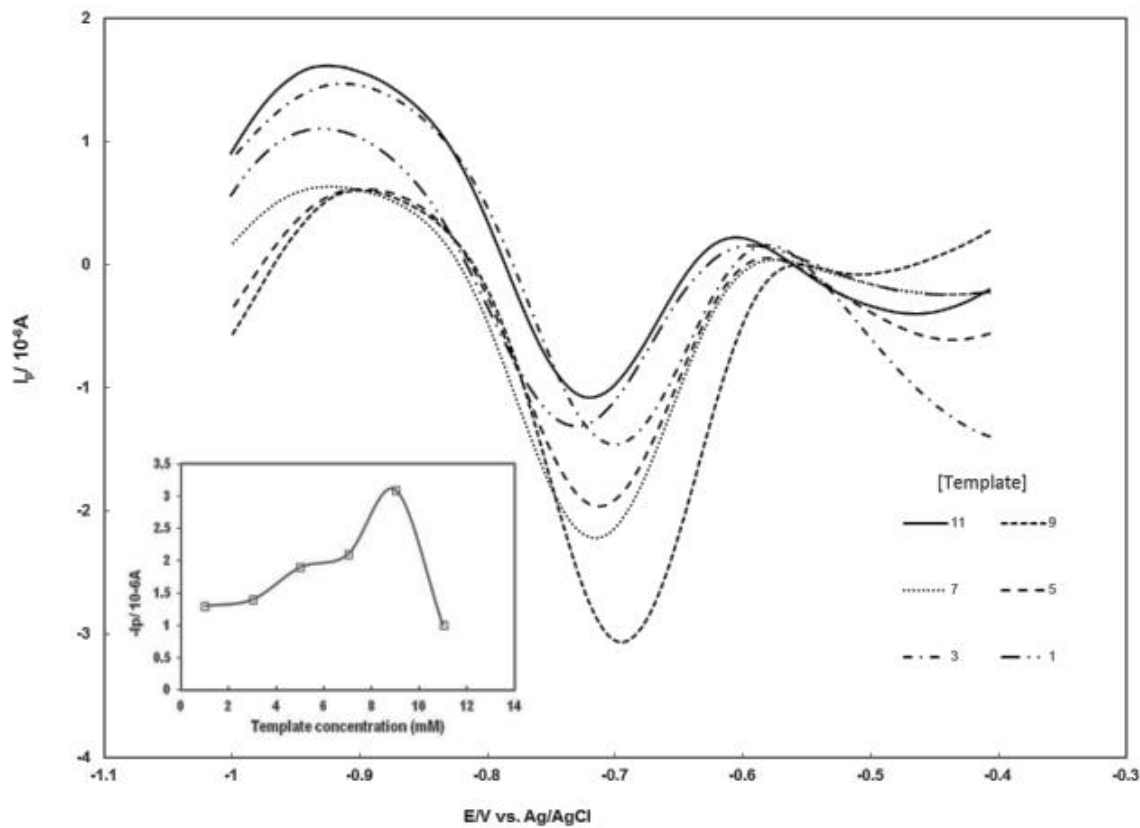

**Fig 4.** Current response diagram of 30 μM PQ at the imprinted Ppy/EBB/GPE in phosphate buffer at pH 7.0 with 0.2 M KCl for different template concentrations

*3.4.2. Interference study*

The selectivity seems to be regulated by the number and entity of the interactions between the target molecules and the imprinted sites into the film. These sites are formed somehow susceptible to be permeable only for those molecules. In this study, most potential interferences with PQ, namely heavy metals, the most abundant ions in natural water, were investigated. This study would reveal that how selective the proposed sensor could be and to what extent it could distinguish target molecules among all concomitants. To that aim, an known amount of probable interferences such as $Fe^{2+}$, $Fe^{3+}$, $Al^{3+}$, $Ca^{2+}$, $Mg^{2+}$, $Cd^{2+}$ and $Co^{2+}$ cations were added to the constant concentration of PQ (30 μM) proportional to the concentration ratios of 1:1, 1:100 and 1:1000 (v/v) (PQ/concomitant). The analytical measurements were conducted under optimized conditions, however, a negligible reduction observed in the current response (<5%) at the peak 0.7 V in all ratios. Consequently, this imprinted sensor was qualified to be engaged to detect PQ in complex samples.

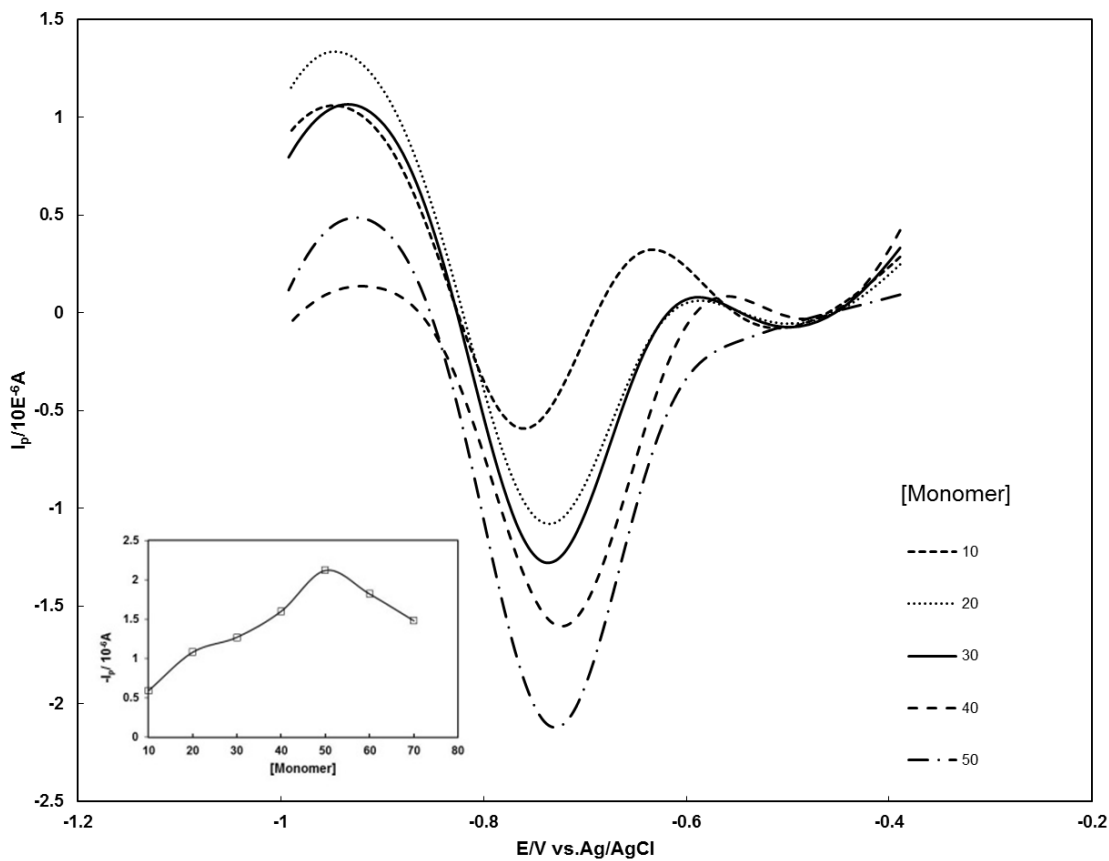

**Fig 5.** Differential pulse voltammogram of 30 μM PQ at the imprinted Ppy/EBB/GPE in phosphate buffer at pH 7.0 with 0.2 M KCl for different monomer concentrations

*3.5. Real water samples analysis:* To assess the applicability of the proposed sensor (imprinted Ppy/EBB/GPE) to voltammetric analysis of a realistic sample, a domestic water sample was used. The sample was collected from Karaj dam water and spiked to three different concentrations of PQ (10 μM, 20 μM, 50 μM). The resultant current responses were measured by the DPV method at a potential range between -0.2 to -1.3 V at the optimized conditions. The results are summarized in table (1). As can be inferred from this table, rational recoveries of spiked PQ obtained can prove the applicability of this modified electrode in the realistic samples.

Table 1. Determination of PQ in dam water samples

| Spiked PQ (μM) | Found PQ (μM) | Recovery (%) | RSD* (%) |
|---|---|---|---|
| 10 | 10.26 ± 0.62 | 102.6 | 2.44 |
| 20 | 18.6 ± 1.32 | 93 | 2.84 |
| 50 | 52 ± 2.48 | 104 | 1.92 |

*Relative standard deviation of three samples

## 4. Conclusions

This study considered the reduction of paraquat (PQ), an herbicide which is still being used unlimitedly in agricultural farms. After optimization of the best experimental conditions, the fabricated imprinted Ppy/EBB/GPE was utilized for the samples collected from the dam water. Results showed that accuracy, sensitivity and repeatability of the proposed method were acceptable enough to be used in real analysis of paraquat with satisfactory results. On the basis of the research undertaken, the electroanalytical methods proved to be a promising approach for the determination of environmental and biological pollutants. Molecularly imprinted polymer-based sensors are more feasible ones in order to measure various elements in complex analytes instead of old chromatographic methods.

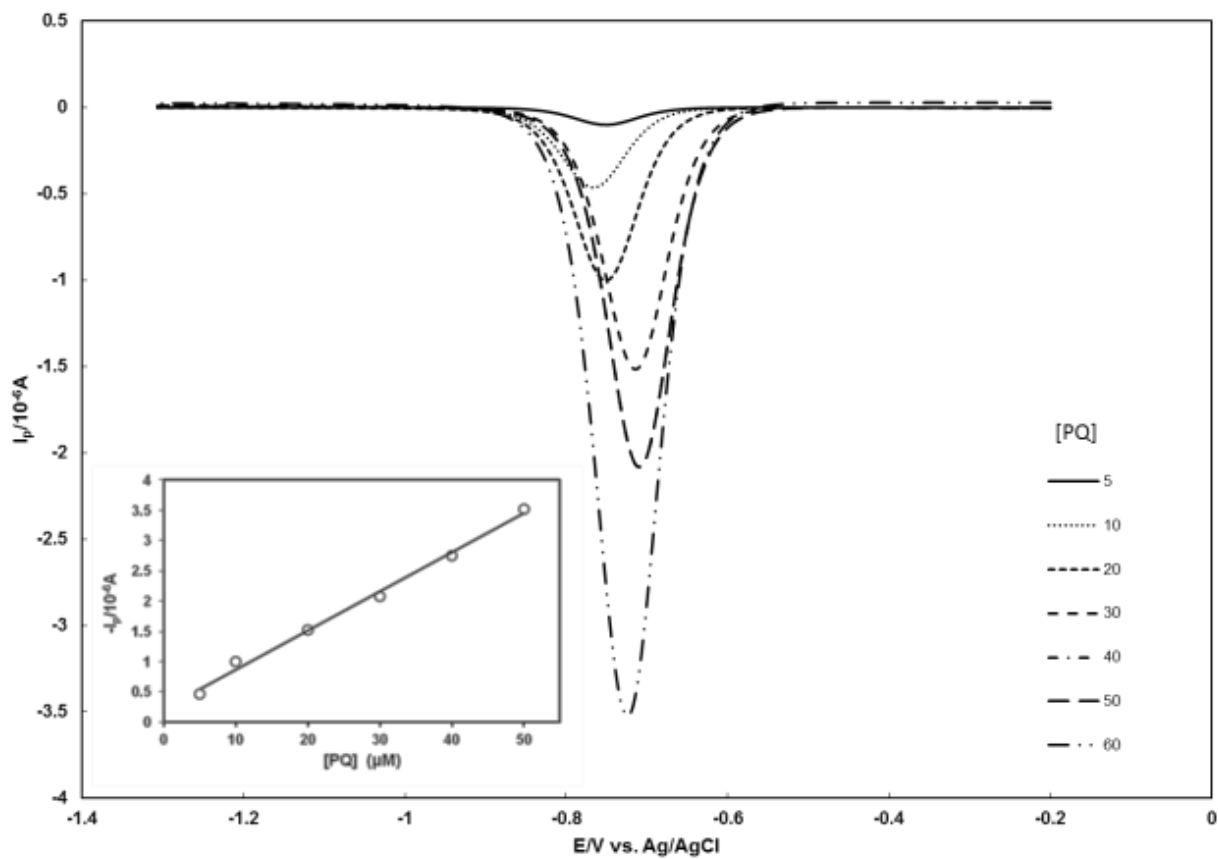

**Fig 6.** Differential pulse voltammogram of 5, 10, 20, 30, 40, 50 μM PQ at imprinted Ppy/EBB/GPE in phosphate buffer at pH 7.0 with 0.2 M KCl

## 5. Acknowledgments